\documentclass[runningheads]{llncs}

\usepackage[T1]{fontenc}
\usepackage{graphicx}
\usepackage{wrapfig}
\usepackage{enumitem}

\usepackage{tikz}
\usetikzlibrary{calc, patterns, arrows, shapes, positioning, fit, cd, tikzmark, decorations.pathreplacing, calligraphy, matrix, decorations.pathmorphing, shapes.multipart}
\usepackage{pgfplots}
\usepackage{hyperref}
\usepackage{color}

\newcommand*{\zenodoDOI}[0]{10.5281/zenodo.10039210}

\usepackage[author=anonymous,marginclue,footnote,draft]{fixme}
\FXRegisterAuthor{mh}{amh}{MH}
\FXRegisterAuthor{sp}{asp}{SP}
\FXRegisterAuthor{dh}{adh}{DH}
\FXRegisterAuthor{ls}{als}{LS}
\FXRegisterAuthor{as}{aas}{AS}

\usepackage{stmaryrd}
\usepackage{amsmath}
\usepackage{amssymb}
\usepackage{xspace}

\usepackage{booktabs, tabularx}

\usepackage{listings}
\usepackage{color}
\definecolor{lightgray}{rgb}{0.9,0.9,0.9}
\spnewtheorem{thm}[theorem]{Theorem}{\bfseries}{\itshape}
\spnewtheorem{cor}[theorem]{Corollary}{\bfseries}{\itshape}
\spnewtheorem{cnj}[theorem]{Conjecture}{\bfseries}{\itshape}
\spnewtheorem{lem}[theorem]{Lemma}{\bfseries}{\itshape}
\spnewtheorem{lemdefn}[theorem]{Lemma and Definition}{\bfseries}{\itshape}
\spnewtheorem{prop}[theorem]{Proposition}{\bfseries}{\itshape}
\spnewtheorem{defn}[theorem]{Definition}{\bfseries}{\upshape}
\spnewtheorem{rem}[theorem]{Remark}{\bfseries}{\upshape}
\spnewtheorem{notation}[theorem]{Notation}{\bfseries}{\upshape}
\spnewtheorem{expl}[theorem]{Example}{\bfseries}{\upshape}
\spnewtheorem{thmdefn}[theorem]{Theorem and Definition}{\bfseries}{\itshape}
\spnewtheorem{propdefn}[theorem]{Proposition and Definition}{\bfseries}{\itshape}
\spnewtheorem{assn}[theorem]{Assumption}{\bfseries}{\upshape}
\spnewtheorem{algorithm}[theorem]{Algorithm}{\bfseries}{\upshape}

\newcommand{\hearts}{\heartsuit}

\newcommand{\tool}{COOL-MC\xspace}
\newcommand{\sem}[1]{\llbracket #1 \rrbracket}
\newcommand{\Agents}[0]{\ensuremath{\mathsf{Ag}}}
\newcommand{\agent}[0]{\ensuremath{a}}

\newcommand{\Vars}[0]{\ensuremath{\mathsf{V}}}
\newcommand{\CLbox}[1]{\ensuremath{[#1]}}
\newcommand{\CLdia}[1]{\ensuremath{\langle#1\rangle}}
\newcommand{\Pow}[0]{\ensuremath{\mathcal{P}}}

\usepackage[capitalise,nameinlink]{cleveref}
\crefname{defn}{Definition}{Definition}

\begin{document}
\title{
  Generic Model Checking for\\ Modal Fixpoint Logics in \tool}
%
%\titlerunning{ Generic Model Checking in \tool}
 \author{
 Daniel Hausmann\orcidID{0000-0002-0935-8602}\inst{1}\thanks{Funded by the ERC Consolidator grant D-SynMA (No.
	772459)} \and
 Merlin Humml\orcidID{0000-0002-2251-8519}\inst{2}\thanks{Funded by the Deutsche Forschungsgemeinschaft (DFG, German Research Foundation) -- project number 377333057 and 393541319/GRK2475/1-2019} \and
 Simon Prucker\orcidID{0009-0000-2317-5565}\inst{2} \and
 Lutz Schröder\orcidID{0000-0002-3146-5906}\inst{2}\thanks{Funded by the Deutsche Forschungsgemeinschaft (DFG, German Research Foundation) -- project number 419850228} \and
 Aaron Strahlberger\inst{2}}
% %
 \authorrunning{D.\ Hausmann, M.\ Humml, S.\ Prucker, L.\ Schröder, A.\ Strahlberger}

\institute{University of Gothenburg, Sweden \and
Friedrich-Alexander-Universität Erlangen-Nürnberg,\\Germany
}

%

%\author{}
%\institute{}
\maketitle              % typeset the header of the contribution

\begin{abstract}
  We report on \tool, a model checking tool for fixpoint logics that
  is parametric in the branching type of models (non-deterministic,
  game-based, probabilistic etc.) and in the next-step modalities used
  in formulae. The tool implements generic model checking algorithms
  developed in coalgebraic logic that are easily adapted to concrete
  instance logics. Apart from the standard modal $\mu$-calculus,
  \tool currently supports alternating-time, graded, probabilistic
  and monotone variants of the $\mu$-calculus, but is also effortlessly
  extensible with new instance logics.  The model checking process is
  realized by polynomial reductions to parity game solving, or,
  alternatively, by a \emph{local} model checking algorithm that
  directly computes the extensions of formulae in a lazy fashion,
  thereby potentially avoiding the construction of the full parity game. We
  evaluate \tool on informative benchmark sets.

\end{abstract}
\section{Introduction}\label{sec:intro}

The $\mu$-calculus~\cite{Kozen83} is one of the most expressive logics
for the temporal verification of concurrent systems. Model checking
the $\mu$-calculus is equivalent to parity game solving, and as such
enjoys diversified tool support in the shape of both well-developed
parity game solving suites such as PGSolver~\cite{friedmannLange009,pgsolvegit} or
Oink~\cite{Dijk18} and dedicated model checking tools such as
mCRL2~\cite{AtifGroote23}. While the $\mu$-calculus is standardly
interpreted over relational transition systems, a wide range of
alternative flavours have emerged whose semantics is variously based
on concurrent games as in the alternating-time
$\mu$-calculus~\cite{AlurEA02}; on probabilistic transition systems as
in the (two-valued) probabilistic
$\mu$-calculus~\cite{CirsteaEA11a,LiuEA15,ChakrabortyKatoen16}; on
counting successors as in the graded
$\mu$-calculus~\cite{KupfermanEA02}; or on neighbourhood structures as
in the monotone $\mu$-calculus, the ambient fixpoint logic of game
logic~\cite{Parikh85,PaulyThesis,EnqvistEA19}. Model checking tools
for such $\mu$-calculi are essentially non-existent or limited to
fragments (see additional comments under `related work'). We present
the generic model checker \tool, which implements generic model
checking algorithms for the \emph{coalgebraic
  $\mu$-calculus}~\cite{CirsteaEA11a} developed in previous
work~\cite{DBLP:conf/concur/HausmannS19}. The coalgebraic
$\mu$-calculus is based on the semantic framework of \emph{coalgebraic
  logic}, which treats systems generically as coalgebras for a set
functor encapsulating the system type, following the paradigm of
\emph{universal coalgebra}~\cite{DBLP:journals/tcs/Rutten00}, and
parametrizes the semantics of modalities using so-called
\emph{predicate liftings}~\cite{Pattinson04,Schroder08}. By fairly
simple instantiation to concrete logics, \tool thus serves as the
first available model checker for the probabilistic $\mu$-calculus,
the graded $\mu$-calculus, and the full alternating-time
$\mu$-calculus AMC (model checking tools for alternating-time temporal
logic ATL, a fragment of the AMC, do exist, as discussed further
below). Besides presenting the tool itself and discussing
implementation issues, we conduct an experimental evaluation of \tool
on benchmark series of parity games~\cite{friedmannLange009,pgsolvegit} that we adapt to
the generalized coalgebraic setting. We thus show that \tool scales
even on series of problems designed to be hard in the relational base
case.

\paragraph*{Related Work} 
As mentioned above, \tool is the only currently available model
checker for most of the logics that it supports, other than the standard
modal $\mu$-calculus (and the main point of its genericity is that
support for further logics can be added easily). We refrain from benchmarking \tool against modal $\mu$-calculus model checkers (e.g. mCRL2~\cite{AtifGroote23}) as this would essentially amount to comparing the respective backend parity game solvers. Model checking tools
for alternating-time temporal logic ATL~\cite{AlurEA02} do exist, such
as MOCHA~\cite{DBLP:conf/icse/AlurAGHKKMMW01},
MCMAS~\cite{DBLP:journals/sttt/LomuscioQR17}, and
UMC4ATL~\cite{unboundedatl2021}, out of which MCMAS appears to be the
fastest one currently available~\cite{unboundedatl2021}. We do compare
\tool to MCMAS on two benchmarks, confirming that MCMAS is faster on
ATL. Note however that ATL model checking works along essentially the
same lines as for CTL, and as such is much simpler than model checking
the alternating-time $\mu$-calculus AMC (e.g., it does not require
parity conditions, and unlike AMC model checking it is known to be
in~\textsc{PTime}~\cite{AlurEA02}), so it is expected that dedicated
ATL model checkers will be faster than an AMC model checker like \tool
on ATL. Local solving has been shown to be advantagous in
model checking for the relational $\mu$-calculus~\cite{StevensStirling98}
and for standard parity
games~\cite{FriedmannLange10}.

\tool uses the basic infrastructure, such as parsers and data
structures for formulae, of the \emph{Coalgebraic Ontology Logic
  Solver (COOL/COOL~2)}~\cite{DBLP:conf/cade/GorinPSWW14,CADE2023}, a
generic reasoner aimed at satisfiability checking rather than model
checking. The algorithms we use
here~\cite{DBLP:conf/concur/HausmannS19} improve on either the
theoretical complexity or the complexity analysis of previous model
checking algorithms for concrete instance logics including the
alternating-time~\cite{AlurEA02}, graded~\cite{FerranteEA08}, and
monotone~\cite{HansenEA17} $\mu$-calculi, as well as of a previous
generic model checking algorithm for the coalgebraic
$\mu$-calculus~\cite{HasuoEA16};
see~\cite{DBLP:conf/concur/HausmannS19} for details.

% Strahlberger~\cite{Aaron} extended the COalgebraic Ontology Logic solver~(COOL) with the model checking capabilities based on the procedures for coalgebraic modal logic developed in Hausmann and Schröder~\cite{DBLP:conf/concur/HausmannS19}.
% As part of that work, CL model checking based on concurrent game frames was implemented serving as a baseline for the current work.

\section{Model Checking for the Coalgebraic $\mu$-Calculus}\label{sec:mc}

\noindent We briefly recall the syntax and semantics of the the
underlying generic logic of \tool, the \emph{coalgebraic
  $\mu$-calculus}~\cite{CirsteaEA11a}, and subsequently sketch two
different model checking algorithms implemented for this logic in
\tool, a \emph{local} algorithm that directly computes extensions of
formulae, and a more global algorithm that reduces instances of the
model checking problem to parity
games~\cite{DBLP:conf/concur/HausmannS19}.

\paragraph{Syntax.}
Formulae of the \emph{coalgebraic $\mu$-calculus} are given by the
following grammar, parametrized over a choice of countable
sets~$\Lambda$ and $\Vars$ of \emph{modalities} and \emph{(fixpoint)
  variables}, respectively.
\begin{gather*}
    \varphi, \psi ::= \top \mid \bot \mid \varphi \land \psi 
    \mid \varphi \lor \psi \mid \hearts\varphi\mid X\mid
    \nu X.\,\varphi \mid     \mu X.\,\varphi 
\end{gather*}
where $X\in\Vars$; we assume that $\Lambda$ contains, for each
modality $\hearts\in\Lambda$, also the dual $\overline{\hearts}$ (with
$\overline{\overline{\hearts}}=\hearts$).  The logic generalizes the
standard $\mu$-calculus by supporting arbitrary monotone modalities
$\hearts$ in place of $\Diamond$ and $\Box$, assuming that the
semantics of $\hearts$ can be defined in the framework of coalgebraic
logic, recalled below.  
%The fixpoint operators $\nu X$ and $\mu X$
%\emph{bind} the variable~$X$, giving rise to standard notions of
%\emph{free} and \emph{bound} variables; a formula is \emph{closed} if
%it does not contain free variables. 
To ensure monotonicity, the logic
does not contain negation as an explicit operator; however, negation
of closed formulae can, as usual, be defined via negation normal
forms.  %, and \emph{clean} if every
%fixpoint variable in it is bound by at most one fixpoint operator.
Given a formula $\varphi$, we let $|\varphi|$ denote its syntactic
size. The algorithms we use work on the (Fischer-Ladner)
\emph{closure}\/ $\mathsf{cl}(\varphi)$ of $\varphi$, a succinct graph
representation of the respective formula, intuitively obtained from
its syntax tree by identifying occurrences of fixpoint variables with
their binding fixpoint operators; we have
$|\mathsf{cl}(\varphi)|\leq |\varphi|$~\cite{Kozen83}.  The
\emph{alternation-depth} $\mathsf{ad}(\varphi)$ of fixpoint formulae
$\varphi=\eta X.\,\psi$ is defined in the usual way as the number of
dependent alternations between least and greatest fixpoints
in~$\varphi$; for a detailed account, see~\cite{KupkeMV22}.

\paragraph{Semantics.}
The semantics of the coalgebraic $\mu$-calculus is parametrized over
the choice of a set functor $\mathcal{F}$ that encapsulates the
branching type of systems, e.g.\ nondeterministic
($\mathcal{F}X=\mathcal{P}X$, the powerset of~$X$) or probabilistic
($\mathcal{F}X=\mathcal{D}X$, the set of discrete probability
distributions on~$X$). Formulae are then evaluated over
\emph{coalgebras} $(C,\xi:C\to \mathcal{F}C)$ for~$\mathcal{F}$, that
is, over generalized transition systems consisting of a set~$C$ of
states and \emph{transition function}~$\xi$ that associates to each
state~$c$ a collection $\xi(c)\in \mathcal{F}C$ of observations and
successors, structured according to~$\mathcal{F}$. For the most basic
case, we can pick $\mathcal{F}=\mathcal{P}$ to be the powerset
functor, so that $\mathcal{F}$-coalgebras are standard transition
systems, with $\xi(c)\in\mathcal{P}C$ being the set of successor states
of $c$.

The semantics of modalities $\hearts\in\Lambda$ is defined in terms of
so-called \emph{predicate liftings}, that is, functions
$\sem{\hearts}$ that lift predicates $D\subseteq C$ on $C$ to
predicates $\sem{\hearts}(D)\in \Pow(\mathcal{F}C)$ on
$\mathcal{F}C$. A state $c\in C$ in a coalgebra $(C,\xi)$ then
satisfies a formula $\hearts\psi$ if
$\xi(c)\in\sem{\hearts}(\sem{\psi})$ where~$\sem{\psi}$ is the set of
states that satisfy~$\psi$.

This concept instantiates to the standard modalities $\Diamond$ and
$\Box$ over transition systems (that is, over coalgebras
$(C,\xi:C\to\Pow C)$ for the functor $\Pow$) by taking predicate
liftings
\begin{align*}
\sem{\Diamond}(D)&=\{E\in\Pow C\mid E\cap D\neq\emptyset\} &
\sem{\Box}(D)&=\{E\in\Pow C\mid E\subseteq D\}.
\end{align*}
For another example, consider \emph{graded} modalities of the shape
$\langle n \rangle $ and $[ n ] $ (for $n\in\mathbb{N}$), expressing
that more than~$n$ successors or all but at most~$n$ successors,
respectively, satisfy the argument formula.  We interpret such
modalities over \emph{graded transition systems}, in which every
transition from one state to another is equipped with a non-negative
integer \emph{multiplicity}; these are coalgebras
$(C,\xi:C\to \mathcal{G}C)$ for the \emph{multiset
  functor}~$\mathcal{G}$ that maps a set $X$ to the set $\mathcal{G}X$
of finite multisets over~$X$, represented as maps~$X\to\mathbb{N}$
with finite support~\cite{DAgostinoVisser02}. For $\theta:C\to\mathbb{N}$ and $D\subseteq C$,
we put $\theta(D)=\Sigma_{d\in D}\theta(d)$, and interpret
$\langle n \rangle $, $[ n ] $ as the predicate liftings
\begin{align*}
\sem{\langle n \rangle}(D)&=\{\theta\in \mathcal{G}C\mid \theta(D)>n\} &
\sem{[n]}(D)&=\{\theta\in \mathcal{G}C\mid \theta(C\setminus D)\leq n\}.
\end{align*}
Having defined the semantics of single modal steps, we now extend the
semantics to the full logic, introducing the \emph{game-based
  semantics} of the coalgebraic $\mu$-calculus (which is equivalent to
a recursively defined algebraic
semantics~\cite{Venema06,DBLP:conf/concur/HausmannS19,HausmannSchroder22}). To
treat least and greatest fixpoints correctly, this semantics uses
\emph{parity games}, which are infinite-duration games played by two
players~$\exists$ and~$\forall$. A
parity game \(G= (V,V_{\exists}, E,\Omega)\) consists of a set~$V$ of
positions, with positions $V_{\exists}\subseteq V$ owned by~$\exists$
and the others by~$\forall$, a \emph{move} relation
$E\subseteq V\times V$, and a priority function
\(\Omega\colon V\to \mathbb{N}\) that assigns a natural number
$\Omega(v)$ to each position $v\in V$.  A \emph{play} is a path in the
directed graph $(V,E)$ that is either infinite or ends in a node
$v\in V$ with no outgoing moves. Finite plays \(v_0v_1\dots v_n\) are
won by $\exists$ if and only if \(v_n\in V_{\forall}\) (i.e.\ if
$\forall$ is stuck); infinite plays are won by~$\exists$ if and only
if the maximal priority that is visited infinitely often is even.
%\(\max\{p\mid \forall j\in\mathbb{N}.\exists k\geq j. \Omega(v_k) =
%p\}\) is even.  
A \emph{(history-free)} $\exists$-\emph{strategy} is a
partial function \(s\colon V_\exists\rightharpoonup V\) that assigns
moves to $\exists$-nodes.  A play \emph{follows} a strategy \(s\) if
for all \(i\geq 0\) such that \(v_i\in V_{\exists}\),
\(v_{i+1}=s(v_i)\).  An $\exists$-strategy \emph{wins} a node
\(v\in V\) if $\exists$ wins all plays that start at \(v\) and follow
\(s\).\\  % The \emph{winning region} of $\exists$ is the set of all nodes
% for which $\exists$ has a winning strategy.

For the remainder of the paper, we fix a functor $\mathcal{F}$,
an $\mathcal{F}$-coalgebra~$(C,\xi)$, a set~$\Lambda$ of
modalities with associated monotone predicate liftings, and
a formula~$\chi$ (that uses modalities from $\Lambda$); 
further we let $\mathsf{cl}=\mathsf{cl}(\chi)$ denote the 
closure of $\chi$, and put $n:=|\mathsf{cl}|$ and $k:=\mathsf{ad}(\chi)$.

\begin{defn}
  The \emph{model checking game} $G_{(C,\xi),\chi}=(V,V_\exists,E,\Omega)$ is
  the parity game defined by the following table, where game nodes
  $v\in V=V_\exists\cup V_\forall$ are of the shape
  $v=(c,\psi)\in C\times \mathsf{cl}$ or
  $v=(D,\psi)\in \Pow(C)\times \mathsf{cl}$.
\begin{center}
\begin{tabular}{|c|c|c|c|}
\hline
node & owner & set of allowed moves \\
\hline
$(c,\top)$ & $\forall$ & $\emptyset$ \\
$(c,\bot)$ & $\exists$ & $\emptyset$ \\
$(c,\varphi\land\psi)$ & $\forall$ & $\{(c,\varphi),(c,\psi)\}$ \\
$(c,\varphi\lor\psi)$ & $\exists$ & $\{(c,\varphi),(c,\psi)\}$ \\
$(c,\eta X.\,\psi)$ & $\exists$ & $\{(c,\psi[\eta X.\,\psi/X])\}$ \\
$(c,\hearts\psi)$ & $\exists$ & $\{(D,\psi)\mid \xi(c)\in\sem{\hearts}(D)\}$ \\
$(D,\psi)$ & $\forall$ & $\{(d,\psi)\mid d\in D\}$ \\
\hline
\end{tabular}
\end{center}
\end{defn}

\noindent In order to show satisfaction of $\hearts\psi$ at $c\in C$,
player $\exists$ thus has to claim satisfaction of $\psi$ at a
sufficiently large set $D\subseteq C$ of states; player $\forall$ in
turn can challenge the satisfaction of $\psi$ at any node $d\in D$.

As usual in $\mu$-calculi, the priority function $\Omega$ serves to
detect that the outermost fixpoint that is unfolded infinitely
often is a greatest fixpoint. It is thus defined ensuring that for
nodes $(c,\varphi)$, $\Omega(c,\varphi)$ is even if
$\varphi=\nu X.\,\psi$, odd if $\varphi=\mu X.\,\psi$, and
$\Omega(c,\varphi)=0$ otherwise, and moreover that larger numbers are
assigned to outer fixpoints, using the alternation depth of
fixpoints. %For instance for
%\begin{align*}
%\varphi_X&=\mu X.\,\varphi_Y & \varphi_Y& =\nu Y.\,\varphi_Z & \varphi_Z=\mu Z.\,\psi(X,Y,Z),
%\end{align*} 
%we get $\Omega(c,\varphi_X)=3$, $\Omega(c,\varphi_Y)=2$ and
%$\Omega(c,\varphi_Z)=1$ for $c\in C$. 
The formal definition of $\Omega$ follows
the standard method, see e.g.~\cite{KupkeMV22}.

We say that $c\in C$ \emph{satisfies} $\chi$ (denoted $(C,\xi),c\models \chi$)
if and only if player $\exists$ wins the position $(c,\chi)$ in $G_{(C,\xi),\chi}$.
The \emph{model checking problem} for the coalgebraic $\mu$-calculus consists in deciding, for state $c\in C$ in a coalgebra $(C,\xi)$, and formula $\chi$ of the 
coalgebraic $\mu$-calculus, whether $(C,\xi),c\models\chi$.

We point out that $G_{(C,\xi),\chi}$ is a parity game with $k$ priorities
that contains up to $n\cdot 2^{|C|}$ positions of the form $(D,\psi)$ for
$D\subseteq C$. Therefore it is not feasible to perform model checking
by explicitly constructing and solving this parity game. In previous
work~\cite{DBLP:conf/concur/HausmannS19,HausmannSchroder21}, we have shown that the model
checking problem for the coalgebraic $\mu$-calculus is in
$\textsc{NP}\cap\textsc{co-NP}$ and in $\textsc{QP}$ (under mild
assumptions on the complexity of evaluating single modal steps using
the predicate liftings), providing two methods to circumvent the
explicit construction of the full game:
\begin{enumerate}[wide]
\item Compute the winning region in $G_{(C,\xi),\chi}$ as a nested fixpoint over the set
of positions of the shape $(c,\psi)$; intuitively, this avoids the explicit construction of the intermediate positions of the shape $(D,\psi)$ by directly computing the extension of
subformulae over $C$. This solution is generic in the sense that it works for any instance of the coalgebraic $\mu$-calculus.
\item Provide a polynomial-sized game-characterization of the
  modalities of the concrete logic at hand, enabling a polynomial
  reduction of the model checking problem to solving parity
  games. This makes it possible to use
  parity game solvers, but relies on the logic-specific
  game characterization of the modalities.
\end{enumerate}
As part of this work, we have implemented and evaluated both methods as an extension of the
reasoner COOL, as described next.

% \begin{proof}
% TODO: Instantiate proof from \cite{DBLP:conf/concur/HausmannS19} to
% the two functors.
% \end{proof}

\section{Implementation -- Model Checking in \tool}

% We defined the formal semantics of the modal operator to fit our implementation. Originally the all effective sets are up closed and the condition hold for element containment and not set inclusion, but due to memory and performance reasons the implementation of set inclusion is preferred.

We report on the implementation of model checking for the coalgebraic
$\mu$-calculus within the framework provided by the COalgebraic
Ontology Logic solver (COOL), a coalgebraic reasoner for modal
fixpoint logics \cite{DBLP:conf/cade/GorinPSWW14}, implemented in
OCaml.  The satisfiability-checking capacities of COOL have been
reported elsewhere~\cite{CADE2023}. Our tool~\tool extends this
framework with comprehensive functionality for model checking, along
the lines of \cref{sec:mc}.  To this end, we use existing
infrastructure and data structures of COOL for parsing and
representing (the closure of) input formulae $\chi$ for an extensible
selection of logics, induced by the choice of a set functor
$\mathcal{F}$; a newly added parser reads input models
$(C,\xi:C\to\mathcal{F}(C))$ in the form of coalgebras for the
selected functor (more details on the introduced specification 
format for coalgebras can be found in the artifact~\cite{HausmannEA23Artifact}). 
We thus obtain model checking support for
\begin{itemize}
\item the standard modal $\mu$-calculus (including its fragment CTL)~\cite{Kozen83},
\item the monotone $\mu$-calculus (including its fragment game
  logic)~\cite{Parikh85,PaulyThesis,EnqvistEA19},
\item the alternating-time $\mu$-calculus (including its fragment ATL)~\cite{AlurEA02},
\item the graded $\mu$-calculus~\cite{KupfermanEA02},
\item the probabilistic $\mu$-calculus~\cite{CirsteaEA11a,LiuEA15,ChakrabortyKatoen16}.
\end{itemize}
By the relation between $\mu$-calculus model checking and the solution of games
with parity conditions, made more precise in \cref{sec:exp} below,
\tool can also be seen as a generic qualitative solver for (standard, monotone, alternating-time, graded and probabilistic) parity games.

The core model checking functionality is provided by implementations
of the two approaches described in \cref{sec:mc}: On the one
hand, we implement the direct evaluation of formulae in the form of a
generic \emph{local model checking algorithm}; on the other hand, we
also implement a \emph{polynomial reduction to standard parity games}
for each of the logics currently supported. Below, we provide intuitive
explanations of the two algorithms, pointing out concrete implementational details
only where the implementation is not straight-forward.

\paragraph{Local Model Checking.}\label{sec:local-mc}

% For local model checking, the following algorithm is implemented in COOL.
% To decide whether \(x\in\sem\varphi\), initialize \(U = \{(\varphi, x)\}\) and \(G = \emptyset\).
% \begin{enumerate}
% \item Expansion: pick \emph{some} \((\psi, y)\in U\), add it to \(G\) and remove it from \(U\). \\
% Add \(h(\psi, y)\backslash G\) to \(U\).
% \item (optional) Propagation: compute \(\textbf{E}_G\) and/or \(\textbf{A}_G\). If \((\varphi, x)\in \textbf{E}_G\) then return "yes",
% if \((\varphi, x)\in \textbf{A}_G\) return "no"
% \item Loop: if \(U\neq\emptyset\) continue with 1.
% \item Final Check: compute \(E_G\), if \((\varphi, x)\in E_G\) then return "yes" otherwise "no"
% \end{enumerate}

%  The algorithm successively expands the coalgebraic version
% of the model checking game, and through the optional step~2.\ supports
% \emph{local} model checking, i.e.\ may avoid exploring the entire
% model and stay below the worst-case complexity in practice. A drawback
% is that computing the fixpoint in the optional step may be costly.
% For the calculation of the fixpoints \(\textbf{A}_G\)
% resp. \(\textbf{E}_G\) in steps~2.\ and~4., 

The local model checking algorithm follows the ideas of~\cite{DBLP:conf/concur/HausmannS19}
by
directly encoding the one-step evaluation of formulae $\psi\in\mathsf{cl}$
by means of functions $\mathsf{eval}_\psi:\Pow(C\times\mathsf{cl})\to \Pow(C\times\mathsf{cl})$, corresponding to all moves in the model checking game that evaluate $\psi$
at some state.
For instance, we have
\begin{align*}
\mathsf{eval}_{\varphi\lor\psi}(X)&=\{(c,\varphi\lor\psi)\mid (c,\varphi)\in X\text{ or }(c,\psi)\in X\}\\
\mathsf{eval}_{\hearts\psi}(X)&=\{(c,\hearts\psi)\mid \xi(c)\in\sem{\hearts}(\{d\mid (d,\psi)\in X\})\}
\end{align*} for
$X\subseteq C\times\mathsf{cl}$, and similar functions for the remaining operators; intuitively, $\mathsf{eval}_\psi(X)$ computes the
set of positions in the model checking game that have formula component $\psi$ and are won by player $\exists$, assuming that it is already known that $\exists$ wins all positions in $X$. Crucially, the evaluation function for 
modal operators skips the exploration of the
intermediate nodes $(D,\psi)$ in the model checking game by directly evaluating
the predicate lifting over the set $X$.
%Given sets $D\subseteq C$ and $X\subseteq D\times\mathsf{cl}$, we define a monotone function 
%$\mathsf{CPre}_D:\Pow(D\times\mathsf{cl})\to \Pow(D\times\mathsf{cl})$ 
%that encodes the evaluation of the one logical operator over the set $X$ by 
%\begin{align*}
%\mathsf{CPre}_D(X)=\,&V\times\{\top\}\,\cup\\
%& \{(c,\varphi\land\psi)\mid (c,\varphi)\in X\text{ and }(c,\psi)\in X\}\,\cup\\
%& \{(c,\varphi\lor\psi)\mid (c,\varphi)\in X\text{ or }(c,\psi)\in X\}\,\cup\\
%& \{(c,\eta X.\,\psi)\mid (c,\psi[\eta X.\,\psi/X])\in X\}\,\cup\\
%& \{(c,\hearts\psi)\mid \xi(c)\in\sem{\hearts}(X_\psi)\},
%\end{align*}
%where $X_\psi=\{(d,\varphi)\in X\mid \varphi=\psi\}$. 
Then we can compute the winning regions in the model checking game as nested fixpoints
of the one-step solving function:
\begin{align*}
\mathsf{win}_\exists &= \mu X_k.\,\nu X_{k-1}.\,\ldots.\,\nu X_0. \cup_{\psi\in\mathsf{cl}}\mathsf{eval}_\psi(X_{\Omega(\psi)})\\
\mathsf{win}_\forall &= \nu X_k.\,\nu X_{k-1}.\,\ldots.\,\mu X_0. \cup_{\psi\in\mathsf{cl}}{\mathsf{eval}_{\neg\psi}}(X_{\Omega(\psi)}),
\end{align*}
assuming w.l.o.g. that $k$ is odd, and denoting by $\Omega(\psi)$ the
priority of all game nodes of the shape $(c,\psi)$;
thus the functions below the fixpoints directly 
correspond to the functions $f$, $g$ from~\cite{DBLP:conf/concur/HausmannS19}, Definition 5, noting that $\Omega(\psi)=0$
whenever $\psi$ is not a fixpoint formula.
We implement this game solving procedure
% the one-step evaluation
% function is implemented as a filter function which selects nodes from
% \(U\subseteq W\times \textsf{cl}(\varphi)\).  It is again implemented
% as
by a higher order function that receives the semantic function for
modalities as an argument, and then computes
the relevant fixpoints by Kleene fixpoint iteration.

The overall local model checking implementation then builds the model checking
game step by step, starting from the initial position $(c,\varphi)$ and adding nodes to
which the respective player can move; crucially, the evaluation functions for
modalities allow us to skip all nodes
of the form $(D,\psi)$ during the exploration of the game arena.
At any point during the game construction, the algorithm can attempt to solve
the partially constructed game by computing the fixpoints defined above, allowing it (in some cases) to finish \emph{early}, that
is, before the whole search space has been explored; this constitutes the \emph{local} nature
of the algorithm in the sense that satisfaction of a formula may be proved or refuted
without traversing the whole model.

\paragraph{Parity Game Model Checking.}

Relying on polynomial reductions of modality evaluation
to game fragments~\cite{DBLP:conf/concur/HausmannS19}, we implement
the generation of model checking parity games in \tool by a higher
order function which traverses the input model and formula and
translates all connectives into game nodes as described in
\cref{sec:mc}, interpreting modal operators using a function it
receives as an argument.  The parity game thus constructed then can be
solved using any parity game solver (including an unoptimized native
solver provided by the \tool framework); the current version of
\tool uses PGSolver as external parity game solver
(support for Oink is planned).

The subgames that evaluate individual modalities in this construction
are specific to the logic at hand.
Due to space restrictions, we provide sketches of the reductions for two
central logics here
and refer to~\cite{DBLP:conf/concur/HausmannS19}, Example 15 for full details.
For the standard $\mu$-calculus, we
have modal positions $(c,\Diamond \psi)$ (or $(c,\Box \psi)$), for
which the one-step evaluation games just consist of that single
position controlled by player $\exists$ (or $\forall$), with moves to
all positions $(d,\psi)$ such that $d\in\xi(c)$.  The evaluation games
for, e.g., graded modalities are significantly more involved: For
instance, from a position $(c,\langle n\rangle \psi)$, the game
proceeds in layers, with one layer for each $d\in C$ to which $c$ has
an edge with multiplicity at least $1$. In each layer, player
$\exists$ decides whether or not to include~$d$ in the set of states
that she claims to satisfy~$\psi$; all game positions also contain a
counter that keeps track of the joint multiplicities of all successors
included so far. Player $\exists$ wins the subgame as soon as this counter
exceeds~$n$ but loses when the subgame exits the final layer while the
counter is still below~$n$. Additionally, player $\forall$ can choose,
for any state $d$ that player $\exists$ decides to use, to either
challenge the satisfaction of $\psi$ at $d$ by continuing the model
checking game a position $(d,\psi)$, or accept the choice of $d$ and
proceed to the next layer of the local game, increasing the counter by
the multiplicity of~$d$ as a successor of~$c$.

%\begin{minted}{ocaml}
%type innerNode =
%  ...
%  (* list represents the tuple *)
%  | Coalition of (int list * M.fe * F.hcFormula) 
%  (* contains the list of reachable successor nodes *)
%  | CoalitionEF of M.fe list                     
%  ...
%
%type mnode =
%  | NormalNode of F.formula * M.fe
%  | InitialNode of (F.formula * M.fe) * M.fe
%  | InnerNode of F.formula * M.fe * innerNode
%\end{minted}
%A parity game is represented as a function
%\begin{minted}{ocaml}
%g : mnode ->  int * player * mnode list * string option
%\end{minted}

\section{Experimental Evaluation of the Implementation}\label{sec:exp}

We experimentally evaluate the performance of our two generic model
checking implementations for all logics currently supported. The main
interest in \tool lies in its genericity, which enables it to cover
a wide range of logics not supported by other tools, so comparison to
other tools is mostly omitted for lack of competitors; additional
discussion is provided below.

\paragraph{Generalized parity games.} As we have seen above, model
checking for (coalgebraic) $\mu$-calculi reduces to solving parity
games. Conversely, parity games can also be solved by model checking:
It is well known that player $\exists$ wins a node $v\in V$ in a
parity game with $k$ priorities if and only if $v$ satisfies the
formula
\[ \chi_k := \mu X_{k}.\,\nu X_{k-1} \dots \mu X_1.\, \nu X_0.\,
  \textstyle\bigvee_{0 \le i \le k} \Omega^{-}(i)\land
  ((V_\exists\land\hearts X_i) \lor (V_\forall\land\overline{\hearts}
  X_i))\] where $\Omega^{-}(i)=\{v\in V\mid\Omega(v)=i\}$,
  $\hearts=\Diamond$ and $\overline{\hearts}=\Box$ and 
  $k$ is w.l.o.g. assumed to be odd.  We
exploit this characterization to lift benchmarking problems for
standard parity games to a coalgebraic level of generality: A parity
game is essentially a Kripke structure with propositional atoms for
priorities and player ownership, that is, a coalgebraic model based
(for transitions) on the powerset functor~$\mathcal{P}$. We generalize
this situation by replacing~$\mathcal{P}$ with other
functors~$\mathcal{F}$, and $\Diamond,\Box$ with suitable pairs
$\hearts,\overline{\hearts}$ of dual modalities. In order to win the
resulting generalized game, player $\exists$ then requires a strategy
that picks, at each game node $v$, a set of successors that satisfies
$\hearts$ if $v\in V_\exists$ or $\overline{\hearts}$ if
$v\in V_\forall$; e.g. in the case of standard games player $\exists$
has to pick a single successor at their nodes ($\hearts=\Diamond$),
while they have to allow all successors at nodes belonging to
$\forall$ ($\overline{\hearts}=\Box$). Furthermore, all plays adhering
to such a strategy have to satisfy the parity condition. We then
systematically enrich given standard parity games to supply additional
functor-specific transition structure in a deterministic way to strike a balance between making the games much harder or much easier than the original game while still making use of the added structure; in our leading examples, we
proceed as follows:
%Having, for each node \(w\) in a parity, the set \(succ(v)\) of successors, we generate the following structures in the coalgebra model:
\begin{itemize}[wide]
\item For the monotone $\mu$-calculus, we construct \emph{monotone
    parity games}; concretely, we build \emph{monotone neighbourhood
    structures}~$N$ (e.g.~\cite{Parikh85}; these are coalgebras for
  the \emph{monotone neighbourhood functor}~\cite{HansenKupke04}), in
  which two consecutive steps in the original parity game $G$ are
  merged so that a single step in~$N$ corresponds to the evaluation of
  two-step strategies in~$G$, that is, we define $\xi(v)$ to be the
  set $\{D_1,\ldots,D_m\}$ of (minimal) neighbourhoods
  $D_i\subseteq V$ such that the owner of $v$ has a strategy in $G$ to
  ensure that starting from~$v$ and playing two steps, some node from
  $D_i$ is reached. Then, $\Diamond\varphi$ essentially says
  that~$\exists$ can enforce~$\varphi$ (in two steps), while $\Box\varphi$
  says that~$\exists$ cannot prevent~$\varphi$.
  % \item The \emph{alternating-time $\mu$-calculus} has a complicated
  %   structure where it is not easy to come up with a structure where
  %   the game does not get trivially easy. To have a deterministic
  %   way to produce an outcome function reaching a set of successors
  %   we simply give one player the choice of successor and determine
  %   that player by the remainder of the division of the state number
  %   through the number of agents (three by default).
\item For the graded $\mu$-calculus (\cref{sec:mc}), we
  construct \emph{graded parity games} by equipping moves in $G$ with
  multiplicities summing up to at least~$10$ at each
  node, that is, we assign multiplicity
  $(\xi(v))(u)=\lceil10 \div E(v)\rceil$ to each successor
  $u\in E(v)$
  of $v$ in $G$.
  Then we take \(\hearts = \CLdia{5}, \overline{\hearts}=\CLbox{5}\),
  so to win in the graded parity game, player $\exists$ requires a
  strategy that picks more than five moves, counting multiplicities,
  at~$\exists$ nodes, and all but at most five moves at~$\forall$
  nodes. 
\item For the two-valued probablistic $\mu$-calculus, we construct
  \emph{qualitive stochastic parity games} by imposing a uniform
  distribution on the moves, thus obtaining \emph{probabilistic
    transition systems}, which are coalgebras for the
  \emph{distribution functor} that assigns to a set~$X$ the set of
  (discrete) probability distributions on~$X$.  Then we take
  \(\hearts = \CLdia{\frac{1}{2}},
  \overline{\hearts}=\CLbox{\frac{1}{2}}\) where~$\CLdia{p}$ is read
  ``with probability more than~$p$'', so player $\exists$ wins the
  resulting stochastic parity game if they have a strategy that in
  each~$\exists$-move stays within the winning region with probability
  more than $\frac{1}{2}$, and forces~$\forall$ to stay within
  $\exists$'s winning region with probability at least~$\frac{1}{2}$.
\end{itemize}
We apply the above constructions to various established parity game
benchmarking series, and in each case evaluate the respective variant
of the formula~$\chi_k$, thereby solving the respective monotone,
graded, or probabilistic variants of the game.  Specifically, we use
series of \emph{clique games}, \emph{ladder games}, \emph{Jurdzinski
  games}, \emph{Towers of Hanoi games}, and \emph{language inclusion
  games} generated by the parity game solver
{PGSolver}~\cite{friedmannLange009,pgsolvegit}.
%\begin{center}
%  \begin{tabularx}{1.0\linewidth}{r|XX}
%    \toprule
%    Generator & call & priorities \\
%    \midrule
%    Elevator game & \verb|elevatorvergm {size}| & \(3\)\\
%    Unfair elevator game & \verb|elevatorvergm -u {size}| & \(3\)\\
%    Clique game & \verb|cliquegame {size + 1}| & \(\mathtt{size} + 1\)\\
%    Ladder game & \verb|laddergame {size}| & 2\\
%    Jurdzinski game & \verb|jurdzinskigame {size} {size}| & \(2 \times \mathtt{size}\)\\
%    Towers of Hanoi & \verb|towersofhanoi {size}| & \(2\)\\
%    Language inclusion & \verb|langincl {size} {size}| & \(3\)\\
%    \bottomrule
%  \end{tabularx}
%\end{center}

\paragraph{Lazy games.}
To illustrate the potential advantages of local model checking, we
also devise an experiment in which each game from a series of
generalized parity games (as detailed above) is prepended with a node
owned by player $\exists$ which has one move that leads to the
original game, but also a move to an additional self-looping node with
priority \(0\).  The resulting games all have very small solutions
that can be found by the local solver, while global solving becomes
more and more expensive as the parameters of the game grow.

\paragraph{Modulo game.}
To evaluate the alternating-time $\mu$-calculus~\cite{AlurEA02}
instance of the model checking implementation in \tool, we devise a
series of games, parameterized by a number of agents and a number~$m$ of
moves per agent, but with a fixed number of positions $p_0,\dots,p_9$
marked by propositional atoms of the same name. At~$p_j$, the agents
concurrently each pick a number from the set $\{1,\ldots,m\}$, causing
the game to proceed to position $p_{(h+j)\!\!\mod 10}$ where~$h$ is the
sum of the numbers played.  Given a set $C$ of agents, we evaluate the
formulae
$\varphi_1=\textstyle\bigwedge_{0 \le i < 10}\mu X.\, p_i \vee \CLbox{C}X$
and
$\varphi_2=\nu X.\, \mu Y.\,(X \wedge (p_0 \vee \CLbox{C}Y) \wedge (p_{5} \vee \CLbox{C}Y))$
over the modulo game. Formula~$\varphi_1$ says that the
coalition~\(C\) has a joint strategy to reach any given state
eventually, while~$\varphi_2$ expresses the Büchi property that~\(C\)
can enforce that both~$p_0$ and~$p_5$ are visited infinitely often.
% This strikes a balance between having too few moves available and having so many moves that the implementations cannot handle the model size.

\paragraph{Evaluation setup.}
Our main aim in the evaluation is to show that~\tool scales even on the
benchmark series we use, which are designed to be hard. In the
process, we compare the local model checking method with the reduction
to parity games (\cref{sec:mc}). To solve the parity games
obtained, we use PGSolver's~\cite{friedmannLange009,pgsolvegit} implementation of
Zielonka's recursive algorithm; we expect that practical performance
can be further improved by instead using Oink~\cite{Dijk18} 
as a back-end parity game solver, but leave this issue as future work.

For the standard and monotone $\mu$-calculi, the
reduction to parity games is straightforward, blurring the difference
between model checking and parity game solving. For these logics, we
thus refrain from a comparison between \tool and other existing model
checking tools~\cite{AtifGroote23,Landsaat22}, which would essentially
boil down to a comparison of the respective backend parity game
solvers. On the alternating-time $\mu$-calculus (AMC), we do conduct a
brief comparison with the model checker
MCMAS~\cite{DBLP:journals/sttt/LomuscioQR17}
(further comparison between COOL-MC
and MCMAS can be found in the appendix).
We emphasize that the
meaningfulness of such a comparison is limited, as on the one hand,
MCMAS represents models symbolically while \tool uses an
explicit-state representation, and on the other hand, MCMAS only
supports alternating-time temporal logic ATL (for which
parity-game-based model checking is overkill) while \tool supports the
full AMC. For graded and probabilistic $\mu$-calculi, \tool appears to
be the only existing model checker, so for these logics we evaluate
only the two variants of model checking in \tool; we note that the
Probabilistic Symbolic Model Checker (PRISM)~\cite{KwiatkowskaEA11}
uses a specification language based largely on
PCTL~\cite{HanssonJonsson94}, which is incomparable to the two-valued
probabilistic $\mu$-calculus~\cite{ChakrabortyKatoen16}.

Below, we refer to the different
instantiations of \tool by indexing a logic name with
either $l$ (for local model checking) or $g$ (for model checking by
game reduction); for instance ``graded$_g$'' refers to the variant of
\tool that reduces model checking for the graded
$\mu$-calculus to parity game solving.

%The default solving strategy used in the local solver in \tool adjusts the frequency of solving attempts as the underlying structure grows and the computation costs increase.
%To be precise, \tool counts the nodes that have been added since the last solution attempt and performs a new solution attempt once this count reaches a threshold proportion relative to the overall number of undecided nodes.

%To scale this adaptation with the size of the input, first the size \(\mathtt{max\_size}\) of the fully expanded graph is calculated.
%The threshold then starts at \(5\) nodes as propagation is cheap early on and is increased by \(max(10, \mathtt{max\_size} \div 20\) for the following \(15\) unsuccessful propagation attempts. 

%For the experiments in this work, we use the implementation of
%Zielonka's recursive algorithm that is provided by PGSolver~\cite{pgsolvegit}.

\noindent We measure runtimes as well as the sizes of the graph
structures and games constructed, averaging the values measured in our
experiments over at least five executions, with a timeout of 60 seconds.
All experiments have been executed on a machine with an AMD Ryzen 7 2700 CPU and
32GB of RAM. An artifact containing the source code, evaluation scripts, and
benchmarking sets for all experiments described above is available online~\cite{HausmannEA23Artifact}.

\paragraph{Results and interpretation.}

The runtime results on the generalized parity games experiment are
shown in \cref{fig:ladder,fig:langincl,fig:towers}. The trends
for the different logics and variants of generalized games are
similar. 
For readability, we show the measurements for just three
logics in each case; additional results can be found in the appendix
and in the artifact. 
\vspace{-10pt}

\begin{figure}[htbp]

  \,
  \begin{minipage}[t]{.5\linewidth}
    \centering
    \begin{tikzpicture}
      \begin{semilogyaxis}[
        minor tick num=1,
        xtick distance = 10,
        every axis y label/.style=
        {at={(ticklabel cs:0.5)},rotate=90,anchor=center},
        every axis x label/.style=
        {at={(ticklabel cs:0.5)},anchor=center},
        tiny,
        width=\linewidth,
        height=17em,
        transpose legend,
        legend columns=2,
        legend style={at={(0.5,-0.2)},anchor=north},
        ymode=log,
        log ticks with fixed point,
        xlabel={game size parameter},
        ylabel={runtime (s)},
        xmin=1,
        xmax=80,
        ymin=0,
        ymax=60,
        legend entries={standard$_l$, standard$_g$, 
        %Mono$_l$, Mono$_g$, CL$_l$, CL$_g$,
         probabilistic$_l$, probabilistic$_g$, graded$_l$, graded$_g$}]

        \addplot[color=blue,mark size=1.7pt,mark=x] table [col sep=comma, x=size, y=mean] {benchmarks/parity-mod/results/laddergame-K-COOL.csv};
        \addplot[color=black,mark size=1.7pt,mark=x] table [col sep=comma, x=size, y=mean] {benchmarks/parity-mod/results/laddergame-K-PG.csv};

%        \addplot[color=blue,dashed,mark=x] table [col sep=comma, x=size, y=mean] {benchmarks/parity-mod/results/laddergame-Monotone-COOL.csv};
%        \addplot[color=black,dashed] table [col sep=comma, x=size, y=mean] 
%{benchmarks/parity-mod/results/laddergame-Monotone-PG.csv};

 %       \addplot[color=blue,dotted,mark=x] table [col sep=comma, x=size, y=mean] {benchmarks/parity-mod/results/laddergame-CL-COOL.csv};
%        \addplot[color=black,dotted] table [col sep=comma, x=size, y=mean] {benchmarks/parity-mod/results/laddergame-CL-PG.csv};

        \addplot[color=blue,dotted,mark=triangle*] table [col sep=comma, x=size, y=mean] {benchmarks/parity-mod/results/laddergame-PML-COOL.csv};
        \addplot[color=black,dotted,mark=triangle*] table [col sep=comma, x=size, y=mean] {benchmarks/parity-mod/results/laddergame-PML-PG.csv};

        \addplot[color=blue,mark size=1.2pt,mark=*] table [col sep=comma, x=size, y=mean] {benchmarks/parity-mod/results/laddergame-GML-COOL.csv};
        \addplot[color=black,mark size=1.2pt,mark=*] table [col sep=comma, x=size, y=mean] {benchmarks/parity-mod/results/laddergame-GML-PG.csv};

      \end{semilogyaxis}
    \end{tikzpicture}
    \vspace{-10pt}
    \caption{Ladder games runtimes}\label{fig:ladder}
  \end{minipage}
    \begin{minipage}[t]{.5\linewidth}
    \centering
    \begin{tikzpicture}
      \begin{semilogyaxis}[
        xtick distance = 1,
        every axis y label/.style=
        {at={(ticklabel cs:0.5)},rotate=90,anchor=center},
        every axis x label/.style=
        {at={(ticklabel cs:0.5)},anchor=center},
        tiny,
        width=\linewidth,
        height=17em,
        transpose legend,
        legend columns=2,
        legend style={at={(0.5,-0.2)},anchor=north},
        ymode=log,
        log ticks with fixed point,
        xlabel={game size parameter},
        ylabel={runtime (s)},
        xmin=1,
        xmax=11,
        ymin=0,
        ymax=60,
        legend entries={standard$_l$, standard$_g$, monotone$_l$, monotone$_g$, %CL$_l$, CL$_g$,
         %PML$_l$, PML$_g$, 
         graded$_l$, graded$_g$}]

        \addplot[color=blue,mark size=1.7pt,mark=x] table [col sep=comma, x=size, y=mean] {benchmarks/parity-mod/results/langincl-K-COOL.csv};
        \addplot[color=black,mark size=1.7pt,mark=x] table [col sep=comma, x=size, y=mean] {benchmarks/parity-mod/results/langincl-K-PG.csv};

        \addplot[color=blue,dotted,mark=triangle*] table [col sep=comma, x=size, y=mean] {benchmarks/parity-mod/results/langincl-Monotone-COOL.csv};
        \addplot[color=black,dotted,mark=triangle*] table [col sep=comma, x=size, y=mean] {benchmarks/parity-mod/results/langincl-Monotone-PG.csv};

%        \addplot[color=blue,dotted,mark=x] table [col sep=comma, x=size, y=mean] {benchmarks/parity-mod/results/langincl-CL-COOL.csv};
%        \addplot[color=black,dotted] table [col sep=comma, x=size, y=mean] {benchmarks/parity-mod/results/langincl-CL-PG.csv};

%        \addplot[color=blue,dotted,mark=triangle*] table [col sep=comma, x=size, y=mean] {benchmarks/parity-mod/results/langincl-PML-COOL.csv};
%        \addplot[color=black,dotted,mark=triangle*] table [col sep=comma, x=size, y=mean] {benchmarks/parity-mod/results/langincl-PML-PG.csv};

        \addplot[color=blue,mark size=1.2pt,mark=*] table [col sep=comma, x=size, y=mean] {benchmarks/parity-mod/results/langincl-GML-COOL.csv};
        \addplot[color=black,mark size=1.2pt,mark=*] table [col sep=comma, x=size, y=mean] {benchmarks/parity-mod/results/langincl-GML-PG.csv};

      \end{semilogyaxis}
    \end{tikzpicture}
    \vspace{-10pt}

    \caption{Language inclusion game runtime}\label{fig:langincl}
  \end{minipage}
\end{figure}

\vspace{-15pt}

\noindent It appears that the concrete choice of the logic does not
  strongly effect the runtimes of the local solvers (the blue plots in
  \cref{fig:ladder,fig:langincl,fig:towers}).  For game-based
  solving (the black plots), we observe a considerable impact of the
  choice of logic on the runtimes, in particular solving the graded
  and probabilistic parity games through PGSolver takes much longer
  than for the standard variants.  This is in line with expectations:
  As mentioned in the end of \cref{sec:local-mc}, the game
  characterization of the standard (or monotone)
  modalities~$\Diamond,\Box$ is straightforward, but the encoding of
  graded and probabilistic modalities leads to quadratic blow-up in
  the resulting games.  The local solver however, directly evaluates
  modalities and thereby avoids this blow-up so that 
  the performance of the local solver is hardly affected by the concrete
  choice of modalities.

  \noindent On the other hand, game-based solving  typically is faster than local
  solving. We note that the
  native fixpoint computation that \tool uses for local solving is
  completely unoptimized and performs naive Kleene fixpoint iteration,
  while PGSolver is an optimized tool, and in particular its recursive
  algorithm shows good performance in practice. 
\vspace{-20pt}  

\begin{figure}[htbp]
  \begin{minipage}[t]{.5\linewidth}
    \centering
   \begin{tikzpicture}
\pgfplotsset{compat=newest}
      \begin{semilogyaxis}[
        scale only axis, axis y line*=left,
        xtick distance = 1,
        every axis y label/.style=
        {at={(ticklabel cs:0.5)},rotate=90,anchor=center},
        every axis x label/.style=
        {at={(ticklabel cs:0.5)},anchor=center},
        tiny,
        width=14em,
        height=12.2em,
        transpose legend,
        legend columns=2,
        legend style={at={(0.23,-0.28)},anchor=north},
        ymode=log,
        log ticks with fixed point,
        xlabel={game size parameter},
        ylabel={runtime (s)},
        xmin=1,
        xmax=10,
        ymin=0,
        ymax=60,
        legend entries={standard$_g$, prob$_g$, standard$_l$, %Mono$_l$, Mono$_g$, 
        %CL$_l$, CL$_g$,
          prob$_l$
         %,GML$_l$, GML$_g$
         }]
      
        \addplot[color=black,mark size=1.7pt,mark=x] table [col sep=comma, x=size, y=mean] {benchmarks/parity-mod/results/towersofhanoi-K-PG.csv};
        \addplot[color=black,mark=triangle*] table [col sep=comma, x=size, y=mean] {benchmarks/parity-mod/results/towersofhanoi-PML-PG.csv};
        \addplot[color=blue,mark size=1.7pt,mark=x] table [col sep=comma, x=size, y=mean] {benchmarks/parity-mod/results/towersofhanoi-K-COOL.csv};

%        \addplot[color=blue,dotted,mark=triangle*] table [col sep=comma, x=size, y=mean] {benchmarks/parity-mod/results/towersofhanoi-Monotone-COOL.csv};
%        \addplot[color=black,dotted,mark=triangle*] table [col sep=comma, x=size, y=mean] {benchmarks/parity-mod/results/towersofhanoi-Monotone-PG.csv};

%        \addplot[color=blue,dotted,mark=x] table [col sep=comma, x=size, y=mean] {benchmarks/parity-mod/results/towersofhanoi-CL-COOL.csv};
%        \addplot[color=black,dotted] table [col sep=comma, x=size, y=mean] {benchmarks/parity-mod/results/towersofhanoi-CL-PG.csv};

        \addplot[color=blue,mark=triangle*] table [col sep=comma, x=size, y=mean] {benchmarks/parity-mod/results/towersofhanoi-PML-COOL.csv};

%        \addplot[color=blue,mark size=1.2pt,mark=*] table [col sep=comma, x=size, y=mean] {benchmarks/parity-mod/results/towersofhanoi-GML-COOL.csv};
%        \addplot[color=black,mark size=1.2pt,mark=*] table [col sep=comma, x=size, y=mean] {benchmarks/parity-mod/results/towersofhanoi-GML-PG.csv};
      \end{semilogyaxis}

      \begin{axis}[
        scale only axis, axis y line*=right, axis x line=none,
        ytick distance = 0.2,
%        every axis y label/.style=
%        {at={(ticklabel cs:0.5)},rotate=90,anchor=center},
%        every axis x label/.style=
%        {at={(ticklabel cs:0.5)},anchor=center},
        tiny,
        width=14em,
        height=12.2em,
        transpose legend,
        legend columns=2,
       legend style={at={(0.9,-0.28)},anchor=north},
%        ymode=log,
%        log ticks with fixed point,
%        xlabel={game size parameter},
        ylabel={exploration quotient},
        xmin=1,
        xmax=10,
        ymin=0,
        ymax=1,
        legend entries={eq standard$_l$, %Mono$_l$, Mono$_g$, 
        %CL$_l$, CL$_g$,
         eq prob$_l$,
         %,GML$_l$, GML$_g$
         }]
    \addplot[color=red,mark size=1.7pt,mark=x] 
coordinates {(1,1)(2,1)(3,1)(4,1)};
        \addplot[color=red,dotted,mark=triangle*] coordinates {(1,1)(2,0.417989417989)(3,0.227513227513)(4,0.121693121693)
 (5,0.0401724475799)(6,0.0411522633745)(7,0.0414135475864)(8,0.0418635370624)};
;
  \end{axis} 
      
    \end{tikzpicture}
    \vspace{-20pt}
    \caption{Towers of Hanoi runtimes}\label{fig:towers}
  \end{minipage}%
  \,\,\,
    \begin{minipage}[t]{.5\linewidth}
    \centering
    \begin{tikzpicture}
\pgfplotsset{compat=newest}
      \begin{semilogyaxis}[
        scale only axis, axis y line*=left,
        xtick distance = 1,
        every axis y label/.style=
        {at={(ticklabel cs:0.5)},rotate=90,anchor=center},
        every axis x label/.style=
        {at={(ticklabel cs:0.5)},anchor=center},
        tiny,
        width=14em,
        height=12.2em,
        transpose legend,
        legend columns=3,
        legend style={at={(0.23,-0.2)},anchor=north},
        ymode=log,
        log ticks with fixed point,
        xlabel={game size parameter},
        ylabel={runtime (s)},
        xmin=1,
        xmax=10,
        ymin=0,
        ymax=60,
        legend entries={standard$_g$, prob$_g$, graded$_g$, %Mono$_l$, Mono$_g$, 
        %CL$_l$, CL$_g$, 
        standard$_l$, prob$_l$, graded$_l$}]

        \addplot[color=black,mark size=1.7pt,mark=x] table [col sep=comma, x=size, y=mean] {benchmarks/parity-mod/results/easy-hanoi-K-PG.csv};
        \addplot[color=black,dotted,mark=triangle*] table [col sep=comma, x=size, y=mean] {benchmarks/parity-mod/results/easy-hanoi-PML-PG.csv};
        \addplot[color=black,mark size=1.2pt,mark=*] table [col sep=comma, x=size, y=mean] {benchmarks/parity-mod/results/easy-hanoi-GML-PG.csv};
        
        \addplot[color=blue,mark size=1.7pt,mark=x] table [col sep=comma, x=size, y=mean] {benchmarks/parity-mod/results/easy-hanoi-K-COOL.csv};

%        \addplot[color=blue,dashed,mark=x] table [col sep=comma, x=size, y=mean] {benchmarks/results/easy-hanoi-Monotone-COOL.csv};
%        \addplot[color=black,dashed] table [col sep=comma, x=size, y=mean] {benchmarks/results/easy-hanoi-Monotone-PG.csv};

%        \addplot[color=blue,dotted,mark=x] table [col sep=comma, x=size, y=mean] {benchmarks/results/easy-hanoi-CL-COOL.csv};
%        \addplot[color=black,dotted,mark=x] table [col sep=comma, x=size, y=mean] {benchmarks/results/easy-hanoi-CL-PG.csv};

        \addplot[color=blue,dotted,mark=triangle*] table [col sep=comma, x=size, y=mean] {benchmarks/parity-mod/results/easy-hanoi-PML-COOL.csv};

        \addplot[color=blue,mark size=1.2pt,mark=*] table [col sep=comma, x=size, y=mean] {benchmarks/parity-mod/results/easy-hanoi-GML-COOL.csv};

      \end{semilogyaxis}
      \begin{axis}[
        scale only axis, axis y line*=right, axis x line=none,
        ytick distance = 0.2,
%        every axis y label/.style=
%        {at={(ticklabel cs:0.5)},rotate=90,anchor=center},
%        every axis x label/.style=
%        {at={(ticklabel cs:0.5)},anchor=center},
        tiny,
        width=14em,
        height=12.2em,
        transpose legend,
        legend columns=1,
       legend style={at={(0.88,-0.2)},anchor=north},
%        ymode=log,
%        log ticks with fixed point,
%        xlabel={game size parameter},
        ylabel={exploration quotient},
        xmin=1,
        xmax=10,
        ymin=0,
        ymax=1,
        legend entries={eq standard$_l$, %Mono$_l$, Mono$_g$, 
        %CL$_l$, CL$_g$,
         eq prob$_l$,
         eq graded$_l$%, GML$_g$
         }]
    \addplot[color=red,mark size=1.7pt,mark=x] 
coordinates {(1,0.553398058252)(2,0.248908296943)(3,0.0939044481054)(4,0.0327398047099)
 (5,0.0110830254715)(6,0.003713596977)(7,0.00124002001436)(8,0.000413579933392)(9,0.000128210400524)(10,0.0000427395572343)};
        \addplot[color=red,dotted,mark=triangle*] coordinates {(1,1)(2,0.502183406114)(3,0.27018121911)(4,0.132682366456)
 (5,0.0398600038888)(6,0.0420222815819)(7,0.0418126046947)(8,0.0418949216738)};
;
        \addplot[color=red,mark size=1.2pt,mark=*] coordinates {(1,0.990291262136)(2,0.445414847162)(3,0.168039538715)(4,0.0585870189546)
 (5,0.0198327824227)(6,0.00664538406411)(7,0.00221898318359)(8,0.000740090407122)(9,0.000246744544406)(10,0.0000822534875075)};
;
  \end{axis}       
      
    \end{tikzpicture}
    \vspace{-20pt}
    \caption{\emph{Lazy} Towers of Hanoi runtimes}\label{fig:easyhanoi}
  \end{minipage}
\end{figure}
\vspace{-20pt}

  \noindent Also, the generalized
  games used in the benchmarks are constructed from parity games
  designed to be hard to solve; in particular, we observe that with
  the notable exceptions of the language inclusion games
  (\cref{fig:langincl,fig:sizes}) and the probabilistic variant of the
  Towers of Hanoi games (\cref{fig:towers}), these games
  typically do not have small solutions so that the local solver
  cannot play out the strength of on-the-fly model checking.

\vspace{-15pt}
  
\begin{figure}
\begin{center}
\begin{footnotesize}
\begin{tabular}{| l | r | r | r| r| r|}
\hline
Experiment series & parameter & worlds & full graph & lazy graph & game size\\
\hline
\hline
Language incl., monotone & $1$ & $3$ & $93$ & $59$ & $126$ \\
 & $7$ & $313$ & $9,703$ & $937$ & $13,146$\\ 
 & $30$ & $\dagger$ & $\dagger$ & $\dagger$ & $1,099,896$\\ 
\hline
\hline
Lazy Hanoi, standard & $1$ & $5$ & $103$ & $57$ & $133$ \\
 & $5$ & $245$ & $5,143$ & $57$ & $6,613$\\ 
 & $9$ & $19,685$ & $413,383$ & $53$ & $531,493$\\ 
 & $10$ & $59,051$ & $1,240,069$ & $53$ & $\dagger$\\ 
\hline
\hline
Lazy Hanoi, graded & $1$ & $5$ & $103$ & $102$ & $523$\\
& $2$ & $11$ & $229$ & $102$ & $2,345$\\
& $4$ & $83$ & $1,741$ & $102$ & $126,222$\\
& $10$ & $59,051$ & $1,240,069$ & $102$ & $\dagger$\\
\hline
\end{tabular}
\end{footnotesize}
\end{center}
\vspace{-15pt}
\caption{Sizes of (full and lazy) graphs and constructed parity games}\label{fig:sizes}
\end{figure}
  \vspace{-15pt}

\noindent This line of argumentation is substantiated by the lazy games
experiment conducted on games built from the Towers of Hanoi series,
shown in \cref{fig:easyhanoi} (the sizes of the constructed
graphs and games are listed in \cref{fig:sizes}).
These results are representative
for the lazy modifications of the other parity game series as
well. Here, the local solver significantly outperforms the algorithm
that first constructs the full game.  It appears that the local solver
does indeed manage to detect the existence of small winning strategies
in these games, thereby avoiding the full exploration of the search
space.  In each case, the extent to which the local solver explores
the full game is shown in \cref{fig:easyhanoi} with a red plot that depicts the
\emph{exploration quotient}, i.e.\ the percentage of the total number
of nodes that are actually explored.  This effect is observed for all
logics currently supported, including the graded and probabilistic
variants.

\vspace{-15pt}

\begin{figure}[ht]
  \begin{minipage}[t]{.48\linewidth}
    \centering
    \begin{tikzpicture}
      \begin{semilogyaxis}[
        minor tick num=1,
        xtick distance = 1,
        every axis y label/.style=
        {at={(ticklabel cs:0.5)},rotate=90,anchor=center},
        every axis x label/.style=
        {at={(ticklabel cs:0.5)},anchor=center},
        tiny,
        width=\linewidth,
        height=17em,
        transpose legend,
        legend columns=2,
        legend style={at={(0.5,-0.2)},anchor=north},
        ymode=log,
        log ticks with fixed point,
        xlabel={number of moves},
        ylabel={runtime (s)},
        xmin=2,
        xmax=10,
        ymin=0,
        ymax=60,
        legend entries={MCMAS$^2$,MCMAS$^4$, CL$^2_g$, CL$^4_g$, CL$^2_l$, CL$^4_l$}]

        \addplot[color=red,mark size=1.7pt,mark=x]  table [col sep=comma, x=parameter_moves, y=mean] {benchmarks/modulos/results/modulosMCMAS-2.csv};
                \addplot[color=red,mark size=1.7pt,mark=o]  table [col sep=comma, x=parameter_moves, y=mean] {benchmarks/modulos/results/modulosMCMAS-4.csv};

        \addplot[color=black,mark size=1.7pt,mark=x] table [col sep=comma, x=parameter_moves, y=mean] {benchmarks/modulos/results/modulos-2-CL-PG.csv};
        \addplot[color=black,mark size=1.7pt,mark=o] table [col sep=comma, x=parameter_moves, y=mean] {benchmarks/modulos/results/modulos-4-CL-PG.csv};

        \addplot[color=blue,mark size=1.7pt,mark=x] table [col sep=comma, x=parameter_moves, y=mean] {benchmarks/modulos/results/modulos-2-CL-COOL.csv};
        \addplot[color=blue,mark size=1.7pt,mark=o] table [col sep=comma, x=parameter_moves, y=mean] {benchmarks/modulos/results/modulos-4-CL-COOL.csv};

      \end{semilogyaxis}
    \end{tikzpicture}
    \vspace{-10pt}
    \caption{Modulo game runtimes ($\varphi_1$)}\label{fig:modulo}
  \end{minipage}%
  \;
  \begin{minipage}[t]{.48\linewidth}
    \centering
    \begin{tikzpicture}
      \begin{semilogyaxis}[
        minor tick num=1,
        xtick distance = 1,
        every axis y label/.style=
        {at={(ticklabel cs:0.5)},rotate=90,anchor=center},
        every axis x label/.style=
        {at={(ticklabel cs:0.5)},anchor=center},
        tiny,
        width=\linewidth,
        height=17em,
        transpose legend,
        legend columns=2,
        legend style={at={(0.5,-0.2)},anchor=north},
        ymode=log,
        log ticks with fixed point,
        xlabel={number of moves},
        ylabel={runtime (s)},
        xmin=2,
        xmax=10,
        ymin=0,
        ymax=60,
        legend entries={CL$^2_g$, CL$^4_g$, CL$^2_l$,  CL$^4_l$}]

        \addplot[color=black,mark size=1.7pt,mark=x] table [col sep=comma, x=parameter_moves, y=mean] {benchmarks/modulos/results/modulos-2-CL-PG-mu.csv};
        \addplot[color=black,mark size=1.7pt,mark=o] table [col sep=comma, x=parameter_moves, y=mean] {benchmarks/modulos/results/modulos-4-CL-PG-mu.csv};
        
        \addplot[color=blue,mark size=1.7pt,mark=x]  table [col sep=comma, x=parameter_moves, y=mean] {benchmarks/modulos/results/modulos-2-CL-COOL-mu.csv};
        \addplot[color=blue,mark size=1.7pt,mark=o]  table [col sep=comma, x=parameter_moves, y=mean] {benchmarks/modulos/results/modulos-4-CL-COOL-mu.csv};
      \end{semilogyaxis}
    \end{tikzpicture}
    \vspace{-10pt}
    \caption{Modulo game runtimes ($\varphi_2$)}\label{fig:modulo-mu}
  \end{minipage}
\end{figure}

\vspace{-20pt}

\noindent \Cref{fig:modulo,fig:modulo-mu} show the runtimes for
$\varphi_1$ and $\varphi_2$ on the modulo game with~2 and~4 agents,
respectively.  We include runtime plots for MCMAS on~$\varphi_1$,
which is expressible in ATL, while~$\phi_2$ is goes beyond ATL and is
thus not handled by \mbox{MCMAS}. As expected, MCMAS is faster on the
fragment that it supports; presumably, this is due partly to the fact
that ATL allows for dedicated model checking algorithms that avoid
parity games and in fact run in polynomial time~\cite{AlurEA02}.

\section{Conclusions and Future Work}

\noindent We have presented and evaluated the generic model checker
\tool, which implements generic model checking algorithms for the
coalgebraic $\mu$-calculus~\cite{DBLP:conf/concur/HausmannS19}, and
has been instantiated to a range of instance logics. In particular,
\tool thus constitutes the first available model checker for the
two-valued probabilistic
$\mu$-calculus~\cite{CirsteaEA11a,LiuEA15,ChakrabortyKatoen16}, the
graded $\mu$-calculus~\cite{KupfermanEA02}, and the full
alternating-time $\mu$-calculus~\cite{AlurEA02} (model checkers for
alternating-time temporal logic
exist~\cite{DBLP:conf/icse/AlurAGHKKMMW01,DBLP:journals/sttt/LomuscioQR17,unboundedatl2021}). 
The benchmarking results suggest the
direct evaluation of modalities in combination with lazy solving
as a setup for coalgebraic model checking that scales well in practice. An
important issue for future work is to develop and implement symbolic
model checking algorithms for the coalgebraic $\mu$-calculus.

% Given that the cost of conversion from concurrent game frames to effectivity frames is the primary reason why one would still use the former semantics for model checking, a clear direction for future work would be to improve the efficiency of the conversion.
% Both the game frame as well as the effectivity frame representation is specified per world and the set of worlds is preserved so an easy optimization step would be converting the worlds in parallel.
% However, parallelization probably needs to be accompanied with further restructuring of the conversion pipeline to prevent an increase in RAM usage leading to crashes.

% Another similar potential for future work would be to do the conversion lazily and convert and cache the model as the respective modalities are encountered by the model checking procedure.
% This should greatly reduce the upfront cost of calculating the effectivity for every coalition at every state.
% This change would however require further changes in the processing architecture of the COOL reasoner and is therefore deferred to future work.

%\subsubsection{Acknowledgements}

\section*{Data-Availability Statement}
All data to reproduce the findings in this paper are available online.
The COOL-MC source code used to compile the artifact is available at tag \texttt{VMCAI-2024} of the COOL git repository~\cite{coolgit}.
Pre-compiled Linux executables as well as a docker container to reproduce the measurements displayed in the figures and tables of this paper are available online~\cite{HausmannEA23Artifact}.

\bibliographystyle{splncs04}
\bibliography{refs}

\providecommand{\noopsort}[1]{}
\begin{thebibliography}{10}
\providecommand{\url}[1]{\texttt{#1}}
\providecommand{\urlprefix}{URL }
\providecommand{\doi}[1]{https://doi.org/#1}

\bibitem{DBLP:conf/icse/AlurAGHKKMMW01}
Alur, R., de~Alfaro, L., Grosu, R., Henzinger, T.A., Kang, M., Kirsch, C.M.,
  Majumdar, R., Mang, F.Y.C., Wang, B.: {JMOCHA:} {A} model checking tool that
  exploits design structure. In: International Conference on Software
  Engineering, {ICSE} 2001. pp. 835--836. {IEEE} Computer Society (2001).
  \doi{10.1109/ICSE.2001.919196}

\bibitem{AlurEA02}
Alur, R., Henzinger, T.A., Kupferman, O.: Alternating-time temporal logic. J.\
  ACM  \textbf{49},  672--713 (2002). \doi{10.1145/585265.585270}

\bibitem{AtifGroote23}
Atif, M., Groote, J.F.: Understanding Behaviour of Distributed Systems Using
  {mCRL2}. Springer (2023). \doi{10.1007/978-3-031-23008-0}

\bibitem{ChakrabortyKatoen16}
Chakraborty, S., Katoen, J.: On the satisfiability of some simple probabilistic
  logics. In: Logic in Computer Science, {LICS} 2016. pp. 56--65. {ACM} (2016).
  \doi{10.1145/2933575.2934526}

\bibitem{CirsteaEA11a}
C{\^{\i}}rstea, C., Kupke, C., Pattinson, D.: {EXPTIME} tableaux for the
  coalgebraic mu-calculus. Log.\ Methods Comput.\ Sci.  \textbf{7}(3) (2011).
  \doi{10.2168/LMCS-7(3:3)2011}

\bibitem{DAgostinoVisser02}
D'Agostino, G., Visser, A.: Finality regained: A coalgebraic study of
  {Scott}-sets and multisets. Arch.\ Math.\ Logic  \textbf{41},  267--298
  (2002). \doi{10.1007/S001530100110}

\bibitem{Dijk18}
van Dijk, T.: Oink: An implementation and evaluation of modern parity game
  solvers. In: Tools and Algorithms for the Construction and, {TACAS} 2018.
  LNCS, vol. 10805, pp. 291--308. Springer (2018).
  \doi{10.1007/978-3-319-89960-2\_16}

\bibitem{EnqvistEA19}
Enqvist, S., Hansen, H.H., Kupke, C., Marti, J., Venema, Y.: Completeness for
  game logic. In: Logic in Computer Science, {LICS} 2019. pp. 1--13. {IEEE}
  (2019). \doi{10.1109/LICS.2019.8785676}

\bibitem{coolgit}
fauprojects: {COOL - The Coalgebraic Ontology Logic Reasoner (git repository)}.
  \url{https://git8.cs.fau.de/software/cool/-/tree/VMCAI-2024}

\bibitem{FerranteEA08}
Ferrante, A., Murano, A., Parente, M.: Enriched {\({\mu}\)}-calculi module
  checking. Log.\ Methods Comput.\ Sci.  \textbf{4}(3) (2008).
  \doi{10.2168/LMCS-4(3:1)2008}

\bibitem{friedmannLange009}
Friedmann, O., Lange, M.: The {PGSolver} collection of parity game solvers.
  Tech. rep., University of Munich (2009)

\bibitem{FriedmannLange10}
Friedmann, O., Lange, M.: Local strategy improvement for parity game solving.
  In: Proceedings First Symposium on Games, Automata, Logic, and Formal
  Verification, {GANDALF} 2010. {EPTCS}, vol.~25, pp. 118--131 (2010).
  \doi{10.4204/EPTCS.25.13}

\bibitem{DBLP:conf/cade/GorinPSWW14}
Gor{\'{\i}}n, D., Pattinson, D., Schr{\"{o}}der, L., Widmann, F., Wi{\ss}mann,
  T.: {COOL} - {A} generic reasoner for coalgebraic hybrid logics (system
  description). In: International Joint Conference on Automated Reasoning,
  {IJCAR} 2014. LNCS, vol.~8562, pp. 396--402. Springer (2014).
  \doi{10.1007/978-3-319-08587-6\_31}

\bibitem{CADE2023}
G{\"o}rlitz, O., Hausmann, D., Humml, M., Pattinson, D., Prucker, S.,
  Schr{\"o}der, L.: {COOL~2} {--} a generic reasoner for modal fixpoint logics
  (system description). In: Automated Deduction, CADE 2023. LNAI, vol. 14132,
  p. 234–247. Springer (2023). \doi{10.1007/978-3-031-38499-8_14}

\bibitem{HansenKupke04}
Hansen, H.H., Kupke, C.: A coalgebraic perspective on monotone modal logic. In:
  Coalgebraic Methods in Computer Science, {CMCS} 2004. ENTCS, vol.~106, pp.
  121--143. Elsevier (2004). \doi{10.1016/j.entcs.2004.02.028}

\bibitem{HansenEA17}
Hansen, H.H., Kupke, C., Marti, J., Venema, Y.: Parity games and automata for
  game logic. In: Dynamic Logic. New Trends and Applications, {DALI} 2017.
  LNCS, vol. 10669, pp. 115--132. Springer (2018).
  \doi{10.1007/978-3-319-73579-5}

\bibitem{HanssonJonsson94}
Hansson, H., Jonsson, B.: A logic for reasoning about time and reliability.
  Formal Aspects Comput.  \textbf{6}(5),  512--535 (1994).
  \doi{10.1007/BF01211866}

\bibitem{HasuoEA16}
Hasuo, I., Shimizu, S., C\^{\i}rstea, C.: Lattice-theoretic progress measures
  and coalgebraic model checking. In: Principles of Programming Languages, POPL
  2016. pp. 718--732. ACM (2016). \doi{10.1145/2837614.2837673}

\bibitem{HausmannEA23Artifact}
Hausmann, D., Humml, M., Prucker, S., Schr\"oder, L., Strahlberger, A.: Generic
  model checking for modal fixpoint logics in {COOL-MC} (artifact). Zenodo
  (2023). \doi{10.5281/zenodo.8332511}

\bibitem{HausmannSchroder21}
Hausmann, D., Schr{\"{o}}der, L.: Quasipolynomial computation of nested
  fixpoints. In: Tools and Algorithms for the Construction and Analysis of
  Systems, {TACAS} 2021. LNCS, vol. 12651, pp. 38--56. Springer (2021).
  \doi{10.1007/978-3-030-72016-2\_3}

\bibitem{HausmannSchroder22}
Hausmann, D., Schr{\"{o}}der, L.: Coalgebraic satisfiability checking for
  arithmetic {\(\mu\)}-calculi. CoRR  \textbf{abs/2212.11055} (2022).
  \doi{10.48550/arXiv.2212.11055}

\bibitem{DBLP:conf/concur/HausmannS19}
Hausmann, D., Schröder, L.: Game-based local model checking for the
  coalgebraic mu-calculus. In: 30th International Conference on Concurrency
  Theory, {CONCUR} 2019. LIPIcs, vol.~140, pp. 35:1--35:16. Schloss Dagstuhl -
  Leibniz-Zentrum für Informatik (8 2019). \doi{10.4230/LIPIcs.CONCUR.2019.35}

\bibitem{unboundedatl2021}
Kański, M., Niewiadomski, A., Kacprzak, M., Penczek, W., Nabiałek, W.:
  Unbounded model checking for {ATL}. Studia Informatica  \textbf{25}(1--2)
  (2021). \doi{10.34739/si.2021.25.01}

\bibitem{Kozen83}
Kozen, D.: Results on the propositional {$\mu$-calculus}. Theor. Comput. Sci.
  \textbf{27},  333--354 (1983). \doi{10.1016/0304-3975(82)90125-6}

\bibitem{KupfermanEA02}
Kupferman, O., Sattler, U., Vardi, M.Y.: The complexity of the graded
  {\(\mathrm{\mu}\)}-calculus. In: Automated Deduction, CADE-18. LNCS,
  vol.~2392, pp. 423--437. Springer (2002). \doi{10.1007/3-540-45620-1\_34}

\bibitem{KupkeMV22}
Kupke, C., Marti, J., Venema, Y.: Size measures and alphabetic equivalence in
  the {\(\mu\)}-calculus. In: Logic in Computer Science, {LICS} 2022. pp.
  18:1--18:13. {ACM} (2022), \url{https://doi.org/10.1145/3531130.3533339}

\bibitem{KwiatkowskaEA11}
Kwiatkowska, M.Z., Norman, G., Parker, D.: {PRISM} 4.0: Verification of
  probabilistic real-time systems. In: Computer Aided Verification, {CAV} 2011.
  LNCS, vol.~6806, pp. 585--591. Springer (2011).
  \doi{10.1007/978-3-642-22110-1\_47}

\bibitem{Landsaat22}
Landsaat, E.: A model checker for game logic via parity games (2022),
  \url{https://fse.studenttheses.ub.rug.nl/28126/}, {BSc} thesis, University of
  Groningen

\bibitem{LiuEA15}
Liu, W., Song, L., Wang, J., Zhang, L.: A simple probabilistic extension of
  modal mu-calculus. In: International Joint Conference on Artificial
  Intelligence, {IJCAI} 2015. pp. 882--888. {AAAI} Press (2015),
  \url{http://ijcai.org/proceedings/2015}

\bibitem{DBLP:journals/sttt/LomuscioQR17}
Lomuscio, A., Qu, H., Raimondi, F.: {MCMAS:} an open-source model checker for
  the verification of multi-agent systems. Int.\ J.\ Softw.\ Tools Technol.\
  Transf.  \textbf{19}(1),  9--30 (2017). \doi{10.1007/s10009-015-0378-x}

\bibitem{Parikh85}
Parikh, R.: The logic of games and its applications. Ann.\ Discr.\ Math.
  \textbf{24},  111--140 (1985). \doi{10.1016/S0304-0208(08)73078-0}

\bibitem{Pattinson04}
Pattinson, D.: Expressive logics for coalgebras via terminal sequence
  induction. Notre Dame J.\ Formal Log.  \textbf{45}(1),  19--33 (2004).
  \doi{10.1305/ndjfl/1094155277}

\bibitem{PaulyThesis}
Pauly, M.: Logic for Social Software. Ph.D. thesis, Universiteit van Amsterdam
  (2001)

\bibitem{DBLP:journals/logcom/PileckiBJ17}
Pilecki, J., Bednarczyk, M.A., Jamroga, W.: {SMC:} synthesis of uniform
  strategies and verification of strategic ability for multi-agent systems. J.\
  Log.\ Comput.  \textbf{27}(7),  1871--1895 (2017).
  \doi{10.1093/logcom/exw032}

\bibitem{DBLP:journals/tcs/Rutten00}
Rutten, J.J.M.M.: Universal coalgebra: a theory of systems. Theor.\ Comput.\
  Sci.  \textbf{249}(1),  3--80 (2000). \doi{10.1016/S0304-3975(00)00056-6}

\bibitem{Schroder08}
Schr{\"{o}}der, L.: Expressivity of coalgebraic modal logic: The limits and
  beyond. Theor.\ Comput.\ Sci.  \textbf{390}(2-3),  230--247 (2008).
  \doi{10.1016/j.tcs.2007.09.023}

\bibitem{StevensStirling98}
Stevens, P., Stirling, C.: Practical model-checking using games. In: Tools and
  Algorithms for Construction and Analysis of Systems, {TACAS} '98. LNCS,
  vol.~1384, pp. 85--101. Springer (1998). \doi{10.1007/BFb0054166}

\bibitem{pgsolvegit}
tcsprojects: {PGSolver (git repository)}.
  \url{https://github.com/tcsprojects/pgsolver}

\bibitem{Venema06}
Venema, Y.: Automata and fixed point logic: {A} coalgebraic perspective. Inf.\
  Comput.  \textbf{204}(4),  637--678 (2006). \doi{10.1016/j.ic.2005.06.003}

\end{thebibliography}

\newpage
\section*{Appendix}

\Cref{fig:clique,fig:jurdzinski} below show the runtimes
for additional experiments on clique games and Jurdzinski games; the results 
in these experiments show the same trends as the results shown and commented on in the main paper.
\begin{figure}[htbp]
  \begin{minipage}[t]{.5\linewidth}
    \centering
    \begin{tikzpicture}
      \begin{semilogyaxis}[
        xtick distance = 1,
        every axis y label/.style=
        {at={(ticklabel cs:0.5)},rotate=90,anchor=center},
        every axis x label/.style=
        {at={(ticklabel cs:0.5)},anchor=center},
        tiny,
        width=\linewidth,
        height=17em,
        transpose legend,
        legend columns=2,
        legend style={at={(0.5,-0.2)},anchor=north},
        ymode=log,
        log ticks with fixed point,
        xlabel={game size parameter},
        ylabel={runtime (s)},
        xmin=1,
        xmax=11,
        ymin=0,
        ymax=60,
        legend entries={standard$_l$, standard$_g$, monotone$_l$, monotone$_g$, 
        %CL$_l$, CL$_g$, 
        graded$_l$, graded$_g$, PML$_l$, PML$_g$}]

        \addplot[color=blue,mark size=1.7pt,mark=x] table [col sep=comma, x=size, y=mean] {benchmarks/parity-mod/results/cliquegame-K-COOL.csv};
        \addplot[color=black, mark size=1.7pt,mark=x] table [col sep=comma, x=size, y=mean] {benchmarks/parity-mod/results/cliquegame-K-PG.csv};

        \addplot[color=blue,dotted,mark=triangle*] table [col sep=comma, x=size, y=mean] {benchmarks/parity-mod/results/cliquegame-Monotone-COOL.csv};
        \addplot[color=black,dotted,mark=triangle*] table [col sep=comma, x=size, y=mean] {benchmarks/parity-mod/results/cliquegame-Monotone-PG.csv};

%        \addplot[color=blue,dashed,mark=square*] table [col sep=comma, x=size, y=mean] {benchmarks/parity-mod/results/cliquegame-CL-COOL.csv};
%        \addplot[color=black,dashed,mark=square*] table [col sep=comma, x=size, y=mean] {benchmarks/parity-mod/results/cliquegame-CL-PG.csv};

        \addplot[color=blue,mark size=1.2pt,mark=*] table [col sep=comma, x=size, y=mean] {benchmarks/parity-mod/results/cliquegame-GML-COOL.csv};
        \addplot[color=black,mark size=1.2pt,mark=*] table [col sep=comma, x=size, y=mean] {benchmarks/parity-mod/results/cliquegame-GML-PG.csv};

%        \addplot[color=blue,mark=+] table [col sep=comma, x=size, y=mean] {benchmarks/results/cliquegame-PML-COOL.csv};
%        \addplot[color=black] table [col sep=comma, x=size, y=mean] {benchmarks/parity-mod/results/cliquegame-PML-PG.csv};
      \end{semilogyaxis}
    \end{tikzpicture}
    \caption{Clique games runtime}\label{fig:clique}
  \end{minipage}%
      \begin{minipage}[t]{.5\linewidth}
    \centering
    \begin{tikzpicture}
      \begin{semilogyaxis}[
        xtick distance = 1,
        every axis y label/.style=
        {at={(ticklabel cs:0.5)},rotate=90,anchor=center},
        every axis x label/.style=
        {at={(ticklabel cs:0.5)},anchor=center},
        tiny,
        width=\linewidth,
        height=17em,
        transpose legend,
        legend columns=2,
        legend style={at={(0.5,-0.2)},anchor=north},
        ymode=log,
        log ticks with fixed point,
        xlabel={game size parameter},
        ylabel={runtime (s)},
        xmin=1,
        xmax=10,
        ymin=0,
        ymax=60,
        legend entries={standard$_l$, standard$_g$, 
        %Mono$_l$, Mono$_g$, 
        %CL$_l$, CL$_g$, 
	    probabilistic$_l$, probabilistic$_g$, graded$_l$, graded$_g$}]

        \addplot[color=blue,mark size=1.7pt,mark=x] table [col sep=comma, x=size, y=mean] {benchmarks/parity-mod/results/jurdzinskigame-K-COOL.csv};
        \addplot[color=black,mark size=1.7pt,mark=x] table [col sep=comma, x=size, y=mean] {benchmarks/parity-mod/results/jurdzinskigame-K-PG.csv};

%        \addplot[color=blue,dashed] table [col sep=comma, x=size, y=mean] {benchmarks/parity-mod/results/jurdzinskigame-Monotone-COOL.csv};
%        \addplot[color=black,dashed] table [col sep=comma, x=size, y=mean] {benchmarks/parity-mod/results/jurdzinskigame-Monotone-PG.csv};

%        \addplot[color=blue,dashed,mark=square*] table [col sep=comma, x=size, y=mean] {benchmarks/parity-mod/results/jurdzinskigame-CL-COOL.csv};
%        \addplot[color=black,dashed,mark=square*] table [col sep=comma, x=size, y=mean] {benchmarks/parity-mod/results/jurdzinskigame-CL-PG.csv};

        \addplot[color=blue,dotted,mark=triangle*] table [col sep=comma, x=size, y=mean] {benchmarks/parity-mod/results/jurdzinskigame-PML-COOL.csv};
        \addplot[color=black,dotted,mark=triangle*] table [col sep=comma, x=size, y=mean] {benchmarks/parity-mod/results/jurdzinskigame-PML-PG.csv};

        \addplot[color=blue,mark size=1.2pt,mark=*] table [col sep=comma, x=size, y=mean] {benchmarks/parity-mod/results/jurdzinskigame-GML-COOL.csv};
        \addplot[color=black,mark size=1.2pt,mark=*] table [col sep=comma, x=size, y=mean] {benchmarks/parity-mod/results/jurdzinskigame-GML-PG.csv};

      \end{semilogyaxis}
    \end{tikzpicture}
    \caption{Jurdzinski games runtime}\label{fig:jurdzinski}
  \end{minipage}%
\end{figure}
We also present an additional benchmark comparing the performance of COOL-MC to the MCMAS model checker.
The \emph{castle game} has been used for benchmarking in previous work
on ATL model
checking~\cite{DBLP:journals/logcom/PileckiBJ17,unboundedatl2021}.
The game is parametrized over the number of castles and the health
points all castles start with.  Each castle has a corresponding knight
that can, in each turn, either be sent out to attack another castle or
stay and defend the castle.  In each turn, all knights decide
concurrently which other castle they want to attack or if they want to
stay at their castle and defend.  A knight who has attacked in one
turn needs to stay and rest in the next turn.  A castle that has its
knight defending it or resting can block one attack.  Each unblocked
attack on a castle reduces that castle's number of health points by
one. When no health points are left, the castle has lost the game and
can no longer attack; this situation is indicated by propositional
atoms $\mathsf{lost}_{a}$, where $a$ is a knight.

For the castle game we check the following AMC formulas 
(which are expressible in ATL as used for the MCMAS benchmarks) for satisfaction in the initial state. The formula
\[\nu X.\, \neg\mathsf{lost}_\agent \wedge \CLbox{\{\agent\}}X\]
expresses that the knight \(\agent\) has a strategy ensuring that her castle never gets destroyed.
We check this formula for each \(\agent \in \Agents\). Moreover, the formula
\[\textstyle\mu X.\, ((\bigwedge_{\agent \in C}\neg\mathsf{lost}_\agent) \wedge (\bigwedge_{\agent \in \Agents\setminus C}\mathsf{lost}_\agent)) \vee \CLbox{C}X\]
expresses that the coalition \(C\) has a joint strategy to ensure that
all other castles are eventually destroyed while none of the allied
castles (belonging to $C$) are destroyed.  We check this formula for
one coalition of each size.

The castle game has the property that almost none of the joint moves
are equivalent, i.e.\ almost all joint moves lead to a different
outcome.  Additionally, the castle game can be specified in MCMAS
using separated local states of the agents.  We chose a
straightforward encoding where each agent has a boolean variable
\lstinline|ready| capturing whether the agent is ready to attack and
an integer variable \lstinline|hp| holds the current number of health
points of the agent.  The atoms \(\mathsf{lost}_\agent\) are evaluated
to true exactly when the \lstinline|hp| variable of \(\agent\) is
\(0\).  The main difficulty of this encoding lies in the specification
of the \lstinline|Evolution|, which encodes the transition function of
agents, as shown in \cref{lst:castles-agent}: The rules of the game
require counting the number of attackers, but MCMAS provides no direct
way to count; hence one case has to be generated for each possible
number of attackers.  Additionally, all the cases have to be disjoint
as MCMAS will pick otherwise some matching case non-deterministically.
So each of these cases has to list all possible combinations of
attackers and non-attackers in a disjunction.

\begin{figure}[htb]
  \begin{lstlisting}
Agent ag1
  Lobsvars = {  };
  Vars: ready : boolean; hp : 0 .. 2; end Vars 
  Actions = { defend, rest, dead, attack_2 }; 
  Protocol:
    ((hp) >= (1) and ready = true) : { defend, attack_2 };
    ((hp) >= (1) and ready = false) : { rest };
    Other : { dead };  end Protocol 
  Evolution:
    hp = (hp) - (1) and ready = false if (Action = attack_2 and
      ((ag2.Action = attack_1 and (hp) >= (1)) or (hp = 1 and
      ag2.Action = attack_1)));
    hp = (hp) - (0) and ready = true if (!(Action = attack_2) and
      ((ag2.Action = attack_1 and (hp) >= (0)) or (hp = 0 and
      ag2.Action = attack_1)));
    ready = false if ((hp) >= (1) and (!(ag2.Action = attack_1) and
      Action = attack_2));
    ready = true if ((hp) >= (1) and (!(ag2.Action = attack_1) and
      !(Action = attack_2))); end Evolution
end Agent
  \end{lstlisting}
\vspace{-20pt}
  \caption{MCMAS encoding of an agent in the two-castle two-health-point game}\label{lst:castles-agent}
\end{figure}
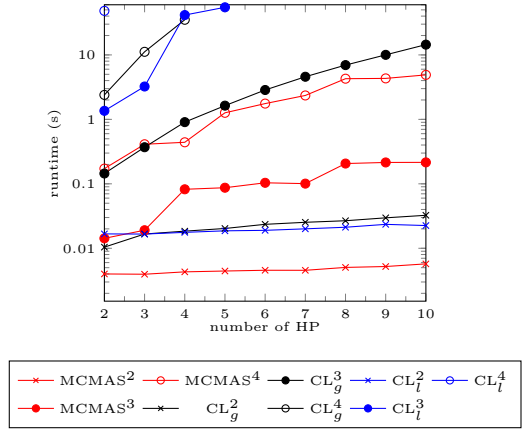
\begin{figure}[htb]
  \begin{minipage}[t]{.48\linewidth}
    \centering
    \begin{tikzpicture}
      \begin{semilogyaxis}[
        minor tick num=1,
        xtick distance = 1,
        every axis y label/.style=
        {at={(ticklabel cs:0.5)},rotate=90,anchor=center},
        every axis x label/.style=
        {at={(ticklabel cs:0.5)},anchor=center},
        tiny,
        width=\linewidth,
        height=17em,
        transpose legend,
        legend columns=2,
        legend style={at={(0.5,-0.2)},anchor=north},
        ymode=log,
        log ticks with fixed point,
        xlabel={number of HP},
        ylabel={runtime (s)},
        xmin=2,
        xmax=10,
        ymin=0,
        ymax=60,
        legend entries={MCMAS$^2$,MCMAS$^3$, MCMAS$^4$, CL$^2_g$, CL$^3_g$, CL$^4_g$, CL$^2_l$, CL$^3_l$, CL$^4_l$}]

        \addplot[color=red,mark size=1.7pt,mark=x]  table [col sep=comma, x=parameter_hp, y=mean] {benchmarks/castles/results/castles-2-MCMAS.csv};
        \addplot[color=red,mark size=1.7pt,mark=*]  table [col sep=comma, x=parameter_hp, y=mean] {benchmarks/castles/results/castles-3-MCMAS.csv};
        \addplot[color=red,mark size=1.7pt,mark=o]  table [col sep=comma, x=parameter_hp, y=mean] {benchmarks/castles/results/castles-4-MCMAS.csv};

        \addplot[color=black,mark size=1.7pt,mark=x] table [col sep=comma, x=parameter_hp, y=mean] {benchmarks/castles/results/castles-2-CL-PG.csv};
        \addplot[color=black,mark size=1.7pt,mark=*] table [col sep=comma, x=parameter_hp, y=mean] {benchmarks/castles/results/castles-3-CL-PG.csv};
        \addplot[color=black,mark size=1.7pt,mark=o] table [col sep=comma, x=parameter_hp, y=mean] {benchmarks/castles/results/castles-4-CL-PG.csv};

        \addplot[color=blue,mark size=1.7pt,mark=x] table [col sep=comma, x=parameter_hp, y=mean] {benchmarks/castles/results/castles-2-CL-COOL.csv};
        \addplot[color=blue,mark size=1.7pt,mark=*] table [col sep=comma, x=parameter_hp, y=mean] {benchmarks/castles/results/castles-3-CL-COOL.csv};
        \addplot[color=blue,mark size=1.7pt,mark=o] table [col sep=comma, x=parameter_hp, y=mean] {benchmarks/castles/results/castles-4-CL-COOL.csv};

      \end{semilogyaxis}
    \end{tikzpicture}
    \caption{Castles game runtimes on the ATL formula series}\label{fig:castles}
  \end{minipage}%
\end{figure}

%
The encoding of the castle game with \(n\) castles and \(h\) health points
in COOL uses \((2 \times H)^n\) as state space where \(2 = \{t, f\}\) and \(H = \{x \mid 0 \le x \le h\}\).
In \cref{fig:castles} we see that COOL-MC again can not match the performance of MCMAS due to the same reasons as mentioned in the paper already.

\end{document}